%% file: Paper2final.tex
\documentclass[11pt,english,twoside]{article}
\pdfoutput=1
\usepackage{jheppub}

\input{mymacros}

\usepackage{graphicx, float, array, xspace, amscd, amsmath, amsthm, amssymb, latexsym, bbm}
\usepackage{arydshln}

\newcommand{\beq}{\begin{equation}}
\newcommand{\eeq}{\end{equation}}

\newcommand{\ii}{\text{i}}
\renewcommand{\ee}{\text{e}}

\newcommand{\ft}[2]{{\textstyle\frac{#1}{#2}}}

\title{Type IIB flux vacua from G-theory II}

\author[a]{Philip Candelas\footnote{candelas@maths.ox.ac.uk}$\!^,$}
\author[b]{Andrei Constantin\footnote{andrei.constantin@physics.uu.se}$\!^,$}
\author[c]{Cesar Damian\footnote{cesaredas@fisica.ugto.mx}$\!^,$}
\author[b]{Magdalena Larfors\footnote{magdalena.larfors@physics.uu.se}$\!^,$}
\author[d]{\linebreak Jose Francisco Morales\footnote{francisco.morales@roma2.infn.it}$\!^,$}

\affiliation[a]{Mathematical Institute, University of Oxford, Andrew Wiles Building, Radcliffe Observatory Quarter, \linebreak Woodstock Road, Oxford, OX2 6GG, UK}
\affiliation[b]{Department of Physics and Astronomy, Uppsala University, SE-751 20, Uppsala, Sweden}
\affiliation[c]{Departamento de Fisica, DCI, Campus Leon, Universidad de Guanajuato, \\C.P. 37150, Leon, Guanajuato, Mexico}
\affiliation[d]{I.N.F.N. Sezione di Roma ``TorVergata'', Dipartimento di Fisica, Universita di Roma ``TorVergata'', \linebreak Via della Ricerca Scientica, 00133 Roma, Italy}

\abstract{We find analytic solutions of type IIB supergravity on geometries that locally take the form
$\text{Mink}\times M_4\times \mathbb{C}$ with $M_4$ a generalised complex manifold. 
The solutions involve the metric, the dilaton, NSNS and RR flux potentials
(oriented along the $M_4$) parametrised by functions varying only over $\mathbb{C}$.  Under this assumption,  the supersymmetry equations
are solved using the formalism of pure spinors in terms of a finite number of holomorphic functions. 
 Alternatively, the solutions can be viewed as vacua of maximally supersymmetric supergravity in six dimensions with a set of scalar fields varying holomorphically over $\mathbb{C}$. 
For a class of solutions characterised by up to five holomorphic functions, we outline how the local solutions can be completed to four-dimensional flux vacua of type~IIB theory. A detailed study of this global completion for solutions with two holomorphic functions has been carried out in the companion paper \cite{paper:global}. The fluxes of the global solutions are, as in F-theory, entirely codified in the geometry of an auxiliary $K3$ fibration over $\mathbb{CP}^1$. The results provide a geometric construction of fluxes in F-theory.}

\preprint{UUITP-19/14} 

\begin{document}
\phantom{}
\vspace{-10pt}
\maketitle

\flushbottom

\newpage
\section{Introduction}

Compactifications of string theories with and without flux is a subject of long history, dating back to the seminal papers \cite{Candelas:1985en,Strominger:1986uh,Hull:1986kz}.\footnote{For recent reviews (with extensive references) on the subject of supersymmetric and non-supersymmetric flux compactifications of string  theory, see \cite{Grana:2005jc,Douglas:2006es,Blumenhagen:2006ci,Koerber:2010bx,Larfors:2013zva}.} Without fluxes, supersymmetry requires that the internal manifold in type~II string compactification is Calabi--Yau, whereas in the presence of fluxes, it must be of generalised Calabi--Yau type. A generalised Calabi-Yau manifold \cite{Hitchin:2004ut} is characterised by the existence of globally defined spinors.  Spinor bilinears define polyforms that behave as
pure spinors in the generalised tangent space. Supersymmetry is preserved in the four-dimensional theory if the pure spinors satisfy  a   
system of first order differential equations~\cite{Grana:2004bg}. If the flux also satisfies the relevant Bianchi identities and the internal manifold is compact, a supersymmetric four-dimensional vacuum is obtained.

In the companion paper \cite{paper:global}, we present concrete examples of supersymmetric four-dimensional 
type~IIB  vacua where all fields can be explicitly written out in an analytic form, even in the presence of fluxes. 
The solutions are built by gluing local solutions on $T^4\times \mathbb{C}$ in a U-duality consistent way. 
Such local solutions can be found by starting from non-compact Calabi-Yau geometries, and then applying a sequence of U-duality transformations that rotate the metric into fluxes. In this way, different classes of flux solutions characterised by up to three holomorphic functions are generated.

The aim of this paper is to present more general flux solutions that cannot be related to  Calabi-Yau geometries by means of U-dualities. We consider geometries that locally take the form $\text{Mink}\times M_4\times \mathbb{C}$ with $M_4$ a generalised complex manifold with $SU(2)$ structure.  We use an ansatz in which the metric, the dilaton and the type IIB fluxes are parametrised by functions varying over the complex plane, and all form potentials are oriented along $M_4$. Under these assumptions, the supersymmetry constraints simplify drastically and can be solved in terms of a finite number of holomorphic functions. We find three classes of solutions with $SU(2)$ structure that we denote A, B and C. The three solutions correspond to different choices of the two angles  describing the relative orientations of the 
two spinors defining the $SU(2)$ structure. The solutions A, B, C in \cite{paper:global} fall into the solution class of that name here for $M_4=T^4$, and correspond to the case where only three of the holomorphic functions characterising the general solutions are non-constant.

  The interest in the solutions  under study here lies in the fact that they can be given an auxiliary, completely geometric description,  following the approach of \cite{Braun:2013yla,Martucci:2012jk}    (related ideas have been explored in 
\cite{Kumar:1996zx,Liu:1997mb,Hellerman:2002ax,Hull:2004in,Flournoy:2004vn,Dabholkar:2005ve,Gray:2005ea,Hull:2007zu,Vegh:2008jn,Pacheco:2008ps, Grana:2008yw,McOrist:2010jw,Andriot:2011uh,Coimbra:2011nw,Berman:2011cg,Berman:2011jh,Coimbra:2011ky,Hohm:2011si,Andriot:2012wx,Andriot:2012an,Blumenhagen:2012nt,Coimbra:2012af, Aldazabal:2013mya, Cederwall:2013naa,Blumenhagen:2013aia,Andriot:2013xca,Cederwall:2014opa}  
%
In particular, solutions  on $T^4$ or $K3$ characterised by $n\leq 5$ holomorphic functions can be extended to the whole complex plane (including infinity) away from a finite number of degeneration points. Around these points, the functions undergo non-trivial monodromy transformations in the U-duality group $SO(2,n,\mathbb{Z})$.  This group is also the modular group of the space of complex structures of an algebraic $K3$ surface with Picard number $20-n$.  Moreover, the locally holomorphic functions parametrise a coset space that is isomorphic to the complex structure moduli space of this $K3$. The $n$ holomorphic functions characterising the flux solution can thus be identified with the $n$ holomorphic parameters (periods of the holomorphic two-form) characterising the complex structure of the $K3$ surface, and the local charges in the flux solution ({\it e.g.}~branes, orientifold planes) can be read off from the monodromy transformations of the periods around singular points in the base. The presence of singularities (and thus local sources) allow non-trivial flux solutions even when the base $\mathbb{C}$ is compactified, in agreement with known no-go theorems \cite{Maldacena:2000mw}.\footnote{Recall that the only globally defined holomorphic function on a compact space is a constant.} Consequently, we can, in this way, construct four-dimensional flux vacua of type IIB string theory in terms of auxiliary geometries that are fibrations of $K3$ surfaces over, for example, a two-sphere. This auxiliary description   is an extension of F-theory \cite{Vafa:1996xn}, in that it provides a geometric description of fluxes in F-theory compactifications. The details of this analysis are given in  \cite{paper:global} for the case $n=2,3$ and will not be repeated here. The techniques developed in that paper can also be applied to $K3$ fibrations with $n>3$ complex parameters, and hence to the local solutions of this paper. Since this computation is very technical, it goes beyond the scope of this paper and is left for future work.

The rest of this note is organised as follows. First, in section \ref{sec:IIb}, we give a very brief review of type IIB supergravity, and present the ansatz we will use for the local solutions. In Section~\ref{sec:local}, we solve the supersymmetry equations and the Bianchi identities (away from local sources) where the  internal six-manifold takes the form $M_4\times \mathbb{C}$. We perform the analysis of the supersymmetry equations using the formalism of pure spinors, briefly reviewed in appendix~\ref{app:purespinors}. We present three classes of solutions, with different flux and metric content, that can be parametrised in terms of a set of holomorphic functions. We also discuss how these different classes of solutions are related by U-duality transformations. 
Finally, in section \ref{sec:conclusion} we draw some conclusions. 
 Appendix \ref{ap:conv} summarises our conventions, and  appendix~\ref{sakilling}
rederives one class of local solutions by the more direct approach of solving the Killing spinor equations.

\section{Type IIB supergravity}
\label{sec:IIb}

In this section, we provide a very brief review of type IIB supergravity, in order to clarify our conventions. For more details, we refer the reader to \cite{Polchinski:1998rr} and recent reviews on flux compactifications \cite{Grana:2005jc,Douglas:2006es,Blumenhagen:2006ci,Koerber:2010bx,Larfors:2013zva}. We also specify the ansatz for the local supersymmetric solutions that will be studied in the next section. 

\subsection{Action and Bianchi identities}
In the low-energy supergravity limit, the bosonic field content of type IIB string theory consists of the  Neveu--Schwarz--Neveu--Schwarz (NSNS) fields (a metric $g$, a scalar field called the dilaton $\phi$ and a two-form field $B$) and the Ramond--Ramond (RR) $p$-form fields $C_p$, where $p$ is $0,2,4$. This is complemented by the fermionic fields: two gravitinos $\Psi_M^A$ and two dilatinos $\lambda^A$, $A=1,2$ of equal chirality. 

The action for the bosonic sector is, in the string frame,
\bea
\label{eq:sIIb}
S &= \frac{1}{2 \kappa_{10} }\int d^{10}x \sqrt{-g} \bigg( e^{-2 \phi} \left[ R + 4 \left( \nabla \phi \right)^2 -\frac{1}{2\cdot 3!} H^2 \right] 
 -\frac{1}{2}F_1^2- \frac{1}{2\cdot3!}F_3^2 -\frac{1}{4\cdot 5!}F_5^2 \bigg) \nonumber \\
&- \frac{1}{4 \kappa_{10}} \int \bigg( C_4 \w H_3 \w F_3 \bigg) ,
\eea
where $g=|{\rm det} \,g_{MN} |$, $H=d B$ and $F_n = d C_{n-1} - H_3 \w C_{n-3}$ are the NSNS and RR field strengths, respectively. In what follows, we collectively refer to the RR fluxes using a polyform language 
\beq
F=d_H C=d C-H\wedge  C~, \mbox{ where } C=C_0+C_2+C_4~.\eeq 
The fluxes must fulfil the Bianchi identities
\beq
d H = 0 \quad  \quad 
d_H F = 0 \; .
\eeq
If sources (NS 5-branes, $D_p$-branes and orientifolds) are present, these will modify the right hand side of these equations.

\subsection{Killing spinor equations}
A purely bosonic supergravity configuration is supersymmetric if and only if the fermionic supersymmetry variations vanish. This leads to the Killing spinor equations (KSE)
\bea \label{eq:Killing_2}
\delta \Psi_M &= \left( 
\nabla_M + \frac{1}{8}\,H_{MNO}\, \Gamma^{NO}\, \cP+\frac{e^\phi}{8}\,\sum\limits_{n}\,\frac{1}{n!}\,F_{P_1 ... P_n}\,\Gamma^{P_1 ... P_n}\, \Gamma_M\, \cP_n
\right)\epsilon =0\\
 \label{eq:Killing_1}
\delta \lambda &= \left(
\Gamma^{M}\,\del_M \phi + \frac{1}{2}\, H_{MNO}\, \Gamma^{MNO}\, \cP- \frac{e^{\phi}}{4}\,\sum\limits_n  \,\frac{(5-n)}{n!}\,F_{P_1 ... P_{n}}\,\Gamma^{P_1 ... P_{n}}\,\cP_n
\right)\epsilon=0
\eea
where $\Psi_M$, $\epsilon$ and $\lambda$  are column vectors containing two Majorana--Weyl spinors of the same chirality, $\nabla$~is the standard covariant derivative, $n$ is odd and $\Gamma$ are the ten-dimensional Dirac matrices (see appendix \ref{ap:conv} for our spinor conventions). The projection matrices $\cP, \cP_n$ are given by 
\beq 
 \cP=-\sigma^3   \qquad \cP_{3}=\sigma^1
 \qquad \cP_{1,5}=\i\,\sigma^2 
\; .
\eeq
where the Pauli matrices $\sigma^i$ are given in appendix \ref{ap:conv}. 

Once the KSE and the Bianchi identities are satisfied in a Minkowski vacuum, it can be shown that all bosonic equations of motion follow \cite{Gauntlett:2005ww,Koerber:2007hd}; as such, the supersymmetric solutions constructed below will satisfy all the constraints required for local type IIB vacua. 

\subsection{The ansatz}
In the following section we will solve the KSE and the Bianchi identities corresponding to type IIB supergravity on space-times $\mathbb{R}^{1,3}\times M_4\times \Sigma$,  with $\Sigma$ an open subset of $\mathbb C$. The torus metric $g_{mn}$, the dilaton $\phi$, the $B$-field and $C_p$-fields are assumed to vary over $\Sigma\subset \mathbb{C}$. All the non-trivial fluxes are assumed to be oriented along $M_4$.
      
Let $\{y^1,y^2,y^3,y^4\}$ be real coordinates on $M_4$ and $z$ a complex coordinate on $\mathbb C$. In these coordinates, we write the metric and the fluxes as:
 \bea
ds^2& =ds_4^2+ds_6^2=e^{2A}\,  \sum_{\mu=0}^3 d x_{\mu} d x^{\mu} +\sum_{m,n=1}^4 g_{mn}\, dy^m\,  dy^n+   e^{2D}\,|h(z)|^2\, d z\, d \bar{z} 
 \label{ans}\\
B &=\frac{1}{2} \,b_{mn}\, dy^m\wedge dy^n \;, \quad     C_2=\frac{1}{2} \,c_{mn}\, dy^m\wedge dy^n \;, \quad C_4=c_4 \, dy^1\wedge dy^2\wedge dy^3 \wedge dy^4 \;,  \nn
\eea
where A, $D$, $g_{mn}, b_{mn}, c_{mn}, c_4, C_0$ and $\phi$ are real $z$-dependent functions.  

To cohere with the ansatz for the bosonic fields, the ten-dimensional Killing spinor $\epsilon= \left(\epsilon^1,\epsilon^2 \right)^T$ must decompose into four- and six-dimensional spinors, that we denote $\zeta_i^a$ and $\eta^i$, respectively. The number of four-dimensional spinors is determined by the number of ten-dimensional spinors and the number $n$ of well-defined internal spinor $\eta^i$:
\beq \label{eq:spindec}
\epsilon^A = \sum_{i=1}^{n} \left( \zeta^A_{i+} \otimes \eta^i_+ +
\zeta^A_{i-} \otimes \eta^i_- \right) \; ,
\eeq
where  $\pm$ denotes chiral and anti-chiral components of the spinors,  $\zeta^A_{i-}=\zeta^{A*}_{i+}$, and $\eta^i_-=\eta^{i*}_+$. We take $\zeta^A_{i\pm}$  to be constant spinors and assume that $\eta^i$ vary only along the $z$-plane, in accordance with our ansatz for the bosonic fields. On $SU(2)$ structure manifolds there are  two globally defined spinors $\eta^1_+$, which can be written in the form (see Appendix \ref{app:purespinors} for details)
\beq
\eta^1_+ =e^{A-\ii \theta\over 2}\,   \eta_+ \quad \quad\quad
\eta^2_+ =e^{A+\ii \theta\over 2 }\,  (\cos \alpha \ \eta_+ +\sin \alpha \ \chi_+)\; ,
\eeq
 Different choices of the angles $\alpha$ and $\theta$ will lead to different kinds of fluxes and brane sources. 

\section{Local supersymmetric solutions}
\label{sec:local} 

In this section, we present three classes of local supersymmetric type IIB solutions, that all satisfy the Killing spinor equations and source-free Bianchi identities. As reviewed in appendix \ref{app:structures},  the fact that $M_4\times \Sigma$ allows two well-defined spinors implies that its structure group is reduced to $SU(2)$. We will use this local $SU(2)$ structure to define pure $O(6,6)$ spinors, and show that these satisfy the supersymmetry equations, once the supergravity fields on $M_4$ vary holomorphically over $\Sigma$.  In appendix \ref{sakilling}, this result is shown without recourse to the pure spinor language for a class of solutions, named type A in the following, with up to five holomorphic functions. 

A six-dimensional manifold of  the local form $M_4 \times \Sigma$ has a local $SU(2)$ structure if it admits a set of differential forms, $(j,\Omega_2)$ on $M_4$ and $K$ on $\Sigma$ ({\it{cf.}} appendix \ref{app:structures}). Choosing local holomorphic coordinates $(z^1,z^2,z)$, we take  for the one-form $K$
  \beq
  \label{eq:K}
  K=e^D \, h(z) \, dz
  \eeq
and expand the two-forms $j,\Omega_2$, the NSNS two-form $B$ and the RR potentials $C_p$ in a basis of closed forms on  $M_4$  with coefficients that depend on $z$. The fluxes of these configurations are given by $H = dB$ and $F_n = d C_{n-1} - H \w C_{n-1}$; they satisfy the Bianchi identities automatically, and always have one leg along either $d z$ or its complex conjugate.

    
    We take the $SU(2)$ structure on $M_4$ to be defined by self-dual two-forms
\beq \label{eq:formdual}
*_4 j = j \quad \quad *_4 \Omega_2 = \Omega_2 \; .
\eeq
   We will consider potential forms oriented along $M_4$. For a $d_4$-closed form $\chi$ on $M_4$ varying only over~$\mathbb{C}$,   the  Hodge dual in six-dimensions can be written as 
 \beq
 * d \chi=*_4 \,  (*_2 d_2 \chi)= -  *_4 \,  d_2^c  \chi     \label{relstard}
 \eeq  
  with  
\beq
  d=\partial +\bar \partial \qquad ~~~~~~~~~~~
   d^c={\rm i} (\bar \partial -\partial)    \label{dc0}
   \eeq
    the exterior derivatives on $ M_4 \times \mathbb{C}$ and $d_2,d_2^c$ their reductions to $\mathbb{C}$.

 The four-dimensional metric will be computed with the help of formula
 \beq
 g_{mn}=-j_{m p}\, I^p{}_n   \label{g4}
 \eeq
with
\beq
I^{p}{}_n=c' \,\epsilon^{ p m_1 m_2 m_3} \, ({\rm Re} \Omega_2)_{n m_1} \, ({\rm Im} \Omega_2)_{m_2 m_3} \label{g42}
\eeq
$\epsilon$ the Levi-Civita symbol in four dimensions, and $c'$ fixed such that $I^2=-1$. These equations follow straightforwardly from the corresponding $SU(3)$ structure identities \eqref{eq:su3I} and \eqref{eq:su3met}, {\it cf.}~Appendix \ref{app:purespinors}.

\subsection{Pure spinor equations}

A particularly elegant reformulation of the supersymmetry constraints  is found using $O(6,6)$ pure spinors (or polyforms)  $\Phi_{1,2}$ \cite{Grana:2004bg,Grana:2005sn}. In terms of these variables the KSE \eqref{eq:Killing_2}-\eqref{eq:Killing_1}  translate into a set of first order differential equations 
\bea \label{eq:pse1}
d_H (e^{3A-\phi} \Phi_1) &= 0 \, \\ \label{eq:pse2}
d_H (e^{2A-\phi} \mbox{Re} \Phi_2) &= 0 \, \\ \label{eq:pse3}
d_H (e^{4A-\phi} \mbox{Im} \Phi_2) &= \frac{e^{4A}}{8} * \lambda(F) \, , 
\eea
with $d_H \chi = d \chi -H \w \chi$ for any differential form $\chi$, and  $ \lambda(F) =  F_1-F_3+ F_5 $. In appendix \ref{app:structures}  we review how $\Phi_{1,2}$ are related to the six-dimensional spinors $\eta^i_\pm$: the latter define nowhere vanishing differential forms $(j, \Omega_2, K)$ which in turn specify  two nowhere vanishing polyforms  $\Phi_{\pm}$  \cite{Jeschek:2004wy,Andriot:2008va}  
\beq
\label{eq:purespin0}
\begin{split}
\Phi_1&= -\frac{ 1}{8}  K \w \left(\sin \alpha \ e^{-\ii j} + \ii \cos \alpha \ \Omega_2 \right) \;  \\[8pt]
\Phi_2 &= \frac{e^{-\ii \theta} }{8}  e^{\frac{1}{2}K\w \bar{K}} \left(\cos \alpha \ e^{-\ii j} - \ii \sin \alpha \ \Omega_2 \right) ~.
\end{split}
\eeq

By specifying $\alpha$ and $\theta$, we will, in the remainder of this section, find three different types of supersymmetric IIB solutions, that we will label A, B and C. More precisely, we will construct local solutions to the KSE and the Bianchi identities following the ansatz \eqref{ans}. In this analysis, we will, from time to time, use the fact that the $SU(2)$ structure defines also an $SU(3)$ structure on the six-dimensional manifold characterised by the forms 
\beq \label{eq:su2su3}
\Omega_3=K \w     \Omega_2   \qquad ~~~~~~~~~~J=j +\frac{\rm i}{2}K\w \bar{K} \; .
\eeq

\subsection{Solution class A} 
\label{sec:typeA} 
We start by considering the case  $\alpha=0$ and $\theta = \pi/2$,  i.e.
\beq
\Phi_1=   -\ft{ {\rm i}}{8}  \, \Omega_3
  \;  \qquad ~~~~\Phi_2  = -\ft{{\rm i} }{8} \,    e^{-{\rm i} J}  
\vspace{-12pt}
\eeq
In this case the two spinors $\eta^A$ are parallel, but out of phase.\footnote{This case was studied in \cite{Martucci:2012jk}, where solutions with the local geometry $K3 \times \Sigma$ were found.} The constraints \eqref{eq:pse1}-\eqref{eq:pse3} become
 \bea  
d_H (e^{3A-\phi} \Omega_3) &= 0 \, ,\\ 
d_H (e^{2A-\phi} \mbox{Im} [e^{- {\rm{i}} J}]) &= 0 \, ,\\ 
d_H (e^{4A-\phi} \mbox{Re} [e^{- {\rm{i}}J}]) &= -e^{4A}* \lambda(F) \, , \label{eq:A3}
\eea
The first two supersymmetry equations lead to
\beq \label{eq:solasusy}
\d(e^{3A-\phi} \Omega_3)= \d(e^{2A-\phi} J) =H \w J =H\wedge \Omega_3= 0  ~,
\eeq
where we solve the first two relations by taking
\beq
\label{eq:su3forms}
\Omega_3=e^{\phi-3A} \hat{\Omega}_3 \qquad J=e^{\phi-2A}\,  \hat{J}   ~,
\eeq
with $d\hat{\Omega}_3=d \hat{J}=0$. A sufficient condition to solve the last two constraints is to take $B$ anti-self-dual, since this implies that $B$ wedges to zero with both $J$ and $\Omega_3$. Since $B$ is a closed $z$-dependent form on $M_4$, this implies that $H = d B$ also wedges to zero with these forms.

The third supersymmetry equation \eqref{eq:A3} then reduces to\footnote{For $p$-forms on even-dimensional spaces, we have $*^2 = -1$ for odd $p$, and $*^2 = +1$ for even $p$.} 
\bea
F_5 &=d C_4- H \wedge C_2=   e^{-4A} *d [e^{4A - \phi}] \\ 
F_3 &= d C_2-H \, C_0= e^{- \phi}  *d B   \\
F_1 &= d C_0=-\ft{1}{2}e^{-4A}*d [e^{4A- \phi} J \w J]
\; .\eea
Using \eqref{eq:hodgewedge}, \eqref{eq:su3star} and \eqref{eq:su3forms}, we compute the Hodge duals 
\begin{equation}
* \frac{1}{2} (d f \w J \w J) 
 = - d^c f \quad \quad
* d f = - \frac{1}{2} d^c f \w J \w J~,
\end{equation}
  On the other hand, using  (\ref{relstard}) and the anti-self-duality\footnote{Notice that for a two-form 
    $*_4 =\hat *_4$, i.e. self-duality with respect to the warped and flat metrics associated to $j$ and~$\hat{j}$ are equivalent.}  of $B$ one finds 
   \beq \label{eq:2dual}
  * d B  = - *_4 d_2^c B =d^c B
   \eeq
    with $d_2$ and $d_2^c$ the restrictions of $d,d^c$ to $\mathbb{C}$.  Consequently
\bea
d C_4- d B \wedge C_2 &=  \ft12 d^c [e^{\phi-4A }\,  \hat{J} \w \hat{J}] \\ 
d C_2-C_0 \, d B &=  e^{-\phi} d^c B  \\
d C_0 &=-d^c e^{- \phi} 
\; .\eea
where we have used \eqref{eq:2dual} and that $B$ is anti-self-dual with respect to $*_4$. 

As noticed in \cite{Martucci:2012jk}, these equations can be written in the compact form
\beq
\label{eq:holpolyform}
\bar \partial  \mathcal{T}=0
\eeq
with the holomorphic polyform
\beq
 \mathcal{T}=e^{-B} (C+{\rm i} e^{-\phi} \mbox{Re} [ e^{- {\rm{i}} j}] ) 
\eeq
or, in components, 
\bea
 \mathcal{T}_0(z)& =C_0 + {\rm i} e^{- \phi}   \\
 \mathcal{T}_2(z) & = C_2 - \tau B  \\
\mathcal{T}_4(z) &= C_4 - B \w C_2 + \ft{1}{2} \tau B \w B - \ft{ \rm{i} }{2} e^{-\phi} j \w j  \; .
\eea
Let us take $B=b_a \chi^{-}_a$ , $C_2=c_a \chi^{-}_a$, $ \mathcal{T}_2 =\beta^{(a)}  \chi^{-}_a$  with $\chi^{-}_a$  a basis of anti-self-dual two-forms on $M_4$ that satisfy 
\beq
\chi^-_a \w \chi^-_b = -2 \delta_{ab} \, d y^1 \w d y^2 \w d y^3\w d y^4\; .
\eeq
A supersymmetric solution is then specified by  the set of holomorphic functions  
\bea
 \tau(z) & =C_0+{\rm i}\, e^{-\phi}   \nn\\
 \beta^{(a)}(z)   &=   c_a- \tau b_a   \nn\\
\sigma(z)  &= -c - 2\, b_a \,  c_a +  \tau\, b_a^2 + \rm{i} \, e^{\phi-4A}  ~,
\label{eq:solAhol}
\eea
where we have used that $j \w j = 2 e^{2\phi-4A} \, d y^1 \w d y^2 \w d y^3\w d y^4$.  We conclude that a supersymmetric solution is specified by $(b_2^- +2)$-holomorphic functions (where $b^-_2$  is the number of
  globally defined anti-self-dual two-forms on $M_4$) characterising the fluxes, and a choice of warped metric for $M_4$.

\subsubsection{Example A: 5 holomorphic functions}
\label{sexamplea}

As an example we can consider $M_4=T^4$ with  trivial $SU(2)$ structure and $K=e^D\, h\, dz$, which give the $SU(3)$ structure forms 
  \bea
  \label{eq:su2sola}
\Omega_3 &=  e^{\phi-3A}\,  h\, dz\wedge   (dy^1+{\rm i}\, dy^4)\wedge   (dy^2+{\rm i} dy^3)   \nn\\
      J &= e^{\phi-2A}   \left[ dy^1\wedge  dy^4 +dy^2\wedge  dy^3 \right]  + \ft{\rm i}{2} \,e^{2D}\, |h|^2\, dz \wedge d\bar z   \; .    
         \eea
For this choice $\ft12 \Omega_2 \wedge \bar \Omega_2=j\wedge j$ implies $D=-A$
  and a basis of anti-selfdual two-forms can be taken to be
  \bea
  \label{eq:chia}
  \chi_a^- & =\{d y^1 \w d y^2 - d y^3 \w d y^4, d y^1 \w d y^3 + d y^2 \w d y^4,d y^1 \w d y^4 - d y^2 \w d y^3  \}\; .
\eea
 The solution is then parametrised by  five holomorphic functions
\[ 
\tau = \tau_1+ \ii \tau_2 \; , \;
\sigma = \sigma_1+ \ii \sigma_2 \; , \;
\beta^{(a)} = \beta^{(a)}_1 + \ii \beta^{(a)}_2 \; , a=1,2,3\; ,
\]
and can be written as 
 \bea
ds^2 &= e^{2 A}\sum_{\mu=0}^3 dx^\mu dx_\mu + e^{\phi - 2 A} \sum_{m,n=1}^4 \delta_{m n} d y^m  d y^n  +  e^{-2 A} | h (z) |^2 dz d\bar{z}  \nn\\
 e^{- \phi } &= \tau_2  , \qquad ~~~~~~~~~~~~~~~~~~ C_0 = \tau_1 , \nn \\
B &=-  \frac{1}{\tau_2} \beta_2^{(a)}  \, \chi_a^{-} 
~, \qquad
~~~~~~~C_2 = \left(\beta_1^{(a)} - \frac{\tau_1}{\tau_2} \beta_2^{(a)} \right) \chi_{a}^{-} , \nn \\
C_4 &=   \left(-\sigma_1+   \frac{2 }{\tau_2} \, \vec \beta_1 \cdot  \vec\beta_2-\frac{ \tau_1}{\tau_2^2}\,  \vec \beta_2 \cdot \vec\beta_2  
\right)
dy^1 \w dy^2 \w dy^3 \w dy^4 ~,
\eea
with $\vec{\beta}_i \cdot \vec \beta_j=\sum_a  \beta^{(a)}_i\, \beta^{(a)}_j $ and
\beq
e^{-2 A} = \sqrt{\sigma_2 \tau_2 - \vec \beta_2 \cdot \vec \beta_2} \; .
\eeq
 The metric has been computed by inserting $J$ and $\Omega_3$ into \eqref{eq:su3I} and \eqref{eq:su3met}.  In appendix \ref{sakilling}, we rederive this  solution by directly solving the equations \eqref{eq:Killing_2}-\eqref{eq:Killing_1} for the ten-dimensional Killing spinors $\epsilon^{1,2}$. 
 
\subsection{Solution class B}
\label{sec:typeB}

In the case $\alpha=\pi/2$, the two spinors $\eta^A$ are orthogonal. The pure spinors are
\beq
\Phi_1 =   -\frac{1}{8} K \,  e^{-\ii j} \quad\quad ~~~~~~~~~~~~~~  \Phi_2 = -\frac{ \rm{i} }{8}  \, \Omega_2 \, e^{\frac{1}{2}   K \w \bar{K}   }  
\eeq
where the $SU(2)$ structure forms $K, j$ and $\Omega_2$ are defined in \eqref{eq:su2def} and we set $\theta=0$ since this phase can be trivially reabsorbed in the definition of $\Omega_2$. 

The supersymmetry equations \eqref{eq:pse1}-\eqref{eq:pse3} then require
\bea  
d_H (e^{3A-\phi} \,K \, e^{-\ii j}) &= 0 \, ,\label{eq:cond0a}\\ 
d_H (e^{2A-\phi} \,\mbox{Im} \Omega_2 ) &= 0 \, ,\label{eq:cond0b}\\ 
d_H (e^{4A-\phi}\, \mbox{Re} \Omega_2  ) &= e^{4A}* \lambda(F) \, , \label{eq:cond0c}
\eea
where we use that $K \w \bar{K}$ is closed and $d \chi \w K \w \bar{K} =0$ for any form $\chi$ that is closed on $M_4$. The first equation implies
\beq \label{eq:cond1}
d (e^{3A-\phi} K) = K \w d (B +\ii j)   =0
\eeq
 that is solved by taking 
\beq
K =e^{\phi-3A}\, h(z) \, d z \qquad d_4 j = 0  \qquad    B + \ii j=\gamma(z)   \label{sol1}
\eeq
with $h(z)$ and $\gamma(z)$ holomorphic zero and two-forms respectively. Equation (\ref{eq:cond0b}) implies
\beq
 d (e^{2A-\phi} \mbox{Im } \Omega_2)=H\wedge  \mbox{Im } \Omega_2= 0 \; .
\eeq
Since $B$ is parallel to $j$ in order to satisfy \eqref{sol1}, the second constraint is automatic. The first may be solved by
\beq
\Omega_2=e^{\phi-2A} \, \hat \Omega_2   
\eeq
with $d \hat \Omega_2=0$. 
 The third supersymmetry equation \eqref{eq:cond0c} decomposes to
\bea
F_1 &=d C_0=0 \\
F_3 &= d (C_2 -B C_0)=e^{-4A} *d (e^{4A-\phi} \mbox{Re} \Omega_2) =  d^c (e^{-2A}  *_4 \mbox{Re} \hat\Omega_2  )    \\
F_5 &= d C_4- H \wedge C_2=0  \; . 
\eea
Using that $\Omega_2$ is self-dual, $*_4 \Omega_2 =  \Omega_2$, this is solved by 
\bea
& -C_2+C_0 B + {\rm i}\, e^{-2A}\,   \mbox{Re} \hat{\Omega}_2  =\rho(z) \\
&  C_4=\ft12 C_0 \, B \wedge B 
\eea
 with $C_0$ a constant and
 $ \rho(z)$ a holomorphic two-form. Writing $B=b_a \chi_a$, $j=j_a \chi_a$, $\gamma=\gamma_a \chi_a$ and
    $C_2-C_0 B =-c\,  \mbox{Re} \hat{\Omega}_2$, the solution is specified by the holomorphic functions 
  \begin{equation}
\rho(z) = c  +{\rm i}\, e^{-2A  }  \quad\quad
\gamma_a(z) = b_a + {\rm i} j_a
\end{equation}
 In the case $M_4=T^4$, after fixing a complex two-form $\hat \Omega_2$, the $j_a$ 
 span a four-dimensional space orthogonal to $\hat \Omega_2$. The dilaton is fixed by the $SU(2)$ condition $j\wedge j =\ft12 \Omega_2 \wedge \bar \Omega_2$.  The solution is then specified by  five holomorphic functions, one from $\rho$ and four from the $\gamma_a$'s. 

\subsubsection{Example B: 4 holomorphic functions}
\label{sexampleb}

As an example of solution in the B-class we can take $M_4=T^4$, $\rho(z)={\rm i}$, {\it i.e.}~$c=A=0$, and
\bea
 \Omega_2 &=e^\phi (dy^1+ {\rm i}\, dy^4)\wedge (dy^2+ {\rm i}\, dy^3) \nn\\
j &= 
\tau_2 dy^1 \w dy^4 + \sigma_2 dy^2 \w dy^3 
-\beta_2^{(1)}(dy^1 \w dy^3 + dy^2 \w dy^4)+\beta_2^{(2)} (dy^1 \w dy^2 - dy^3 \w dy^4 )  \; .   \label{job}
\eea
The condition $j\wedge j =\ft12 \Omega_2 \wedge \bar \Omega_2$ implies
\beq
 e^{2 \phi} =  \sigma_2 \tau_2 -\vec\beta_2  \cdot \vec \beta_2
\eeq
Plugging (\ref{job}) into \eqref{g4}-\eqref{g42} (or the corresponding $J$ and $\Omega_3$ into \eqref{eq:su3I}-\eqref{eq:su3met}) one finds for the metric on $T^4$
\beq
g_{mn} = \left(
 \begin{matrix} 
      \tau_2 & -\beta_{2}^{(1)} & -\beta_{2}^{(2)} & 0 \\
     -\beta_{2}^{(1)} & \sigma_2 & 0 & \beta_{2}^{(2)}\\
      -\beta_{2}^{(2)} & 0 & \sigma_2 & -\beta_{2}^{(1)}\\
      0 &  \beta_{2}^{(2)} & -\beta_{2}^{(1)} & \tau_2 \\
   \end{matrix}
\right)   
\eeq
The solution becomes
\bea
ds^2 &= \sum_{\mu=0}^3 dx^\mu dx_\mu + \sum_{m,n=1}^4 g_{m n} \d y^m  \d y^n  +  e^{2\phi} | h (z) |^2 dz d\bar{z} \; ,  \nn\\
B &=  \tau_1 dy^1 \w dy^4 +\sigma_1 dy^2 \w dy^3- \beta_1^{(1)}(dy^1 \w dy^3 + dy^2 \w dy^4)+\beta_1^{(2)} (dy^1 \w dy^2 - dy^3 \w dy^4 )~,\nn\\
C_0 &= 0 \; , \;   C_2 = 0 \; , \;  C_4 =0   \; .  \label{bsol4}
\eea

\subsection{Solution class C}

\label{sec:typeC}
Finally, we consider the case  $\alpha=0$ and $\theta=\pi$.  Like in case A, the two spinors $\eta^A$ are parallel but now the relative 
phase is simply a sign. The pure spinors are
\beq
\Phi_1 = \frac{1}{8} \Omega_3   \; \; \; \; , \; \; \; \;   \Phi_2 = -\frac{1}{8} e^{- {\rm{i}} J} \; .
\eeq
The supersymmetry equations \eqref{eq:pse1}-\eqref{eq:pse3} thus require
\bea  
d_H (e^{3A-\phi} \Omega_3) &= 0 \, ,\\ 
d_H (e^{2A-\phi} \mbox{Re} [e^{-{ \rm{i}} J}]) &= 0 \, ,\\ 
d_H (e^{4A-\phi} \mbox{Im} [e^{- {\rm{i}} J}]) &= -e^{4A}* \lambda(F) \, .    \label{3eqsC}
\eea
The first two equations imply
\bea
 d (e^{3A-\phi} \Omega_3) = d (e^{2A-\phi} )=H= d J\wedge J=0     \label{djj}
\eea
of which the first three constraints can be solved by taking
  \beq
  \phi=2A    \quad\quad~~~~~ H=0  \quad\quad  ~~~~~\Omega_3=e^{-A } \,   \hat{\Omega}_3  
  \eeq
    with $\d\hat{\Omega}_3 =0$. The six-dimensional manifold is then warped complex  but need not be K\"ahler.
Using $K$ from \eqref{eq:K} in \eqref{eq:su2su3}, we conclude that $\Omega_2=e^{-A-D}\, \hat{\Omega}_2$ with  $\hat{\Omega}_2$
a closed two-form varying holomorphically along the $\mathbb{C}$-plane. 
On the other hand  the last equation in (\ref{djj}) implies
\beq
0 = d J\wedge J= d j \w j - \frac{\rm i}{2} d j \w K \w \bar{K}
 \quad \Leftrightarrow \quad  d j \wedge j=0 ~ \; , ~ d_4 j = 0
 \; ,
\eeq
so $j$ is a closed form on $M_4$ that varies with $z$ in such a way to keep $j \w j$ constant. The equation $d j \wedge j=0$, or equivalently $d^c j \wedge j=0$, can be solved\footnote{For $M_4=T^4$ or $M^4=K3$, which have three-dimensional bases of self-dual two-forms, this is the most general solution, since $d^c j \wedge j=d^c j \wedge\Omega_2=0$ implies that $d^c j$ is anti-selfdual with respect to $*_4$. }  by taking $d^c j$ anti-self-dual with respect to $*_4$
\beq
\label{selfdj}
*_4 d^c j=-d^c j \; .
\eeq
On the other hand, from $j\wedge j =\ft12 \Omega_2 \wedge \bar \Omega_2$ we conclude that
\beq
D=-A    \qquad   \Rightarrow \qquad   \Omega_2=\hat{\Omega}_2 
\eeq 
 Finally the third supersymmetry equation in (\ref{3eqsC}) decomposes into (recall that $H=0$)
\bea
F_1 &= d C_0=0 \\
F_3 &=d C_2= -e^{-4A}  *d (e^{2A} j) =- d^c(e^{-2A}) \wedge j -e^{-2A} * dj  =  -d^c (  e^{-2A} j   )\\
F_5 &=d C_4=0 \; .    \label{eqfc}
\eea
where we used  (\ref{relstard}) and (\ref{selfdj}). 
Eqs. (\ref{eqfc}) can then be solved by taking $C_0, C_4$ constant and 
\be
\gamma=    C_2 + {\rm i}\, e^{-2A}  j  
\eeq
holomorphic.  We recall that $j$ is a two-form  orthogonal to $\Omega_2$ that satisfies  $ d (j \wedge j)=0$. Just as discussed above for solutions of type B, when $M_4=T^4$ we can expand $j$ in a four-dimensional basis of two-forms orthogonal to $\Omega_2$. Writing $C_2=c_a \chi_a$ we build four holomorphic functions 
\beq
\gamma_a= c_{a} + {\rm i}\, e^{-2A}\, j_a  ~. 
\eeq
The flux content of general solutions in this class is then characterised by four holomorphic functions. Additionally, there may be holomorphic functions that parametrise $\Omega_2$.

\subsubsection{Example C: 4 holomorphic functions}
\label{sexamplec}

As an example of solution in the C-class we choose $M_4=T^4$ and
\bea
 \Omega_2 &= (dy^1+ {\rm i}\, dy^4)\wedge (dy^2+ {\rm i}\, dy^3)  \label{joc} \\
 j &= 
e^{2A}\left[ \tau_2 dy^1 \w dy^4 +\sigma_2 dy^2 \w dy^3 
-\beta_2^{(1)}(dy^1 \w dy^3 + dy^2 \w dy^4)+\beta_2^{(2)} (dy^1 \w dy^2 - dy^3 \w dy^4 )  \right]  \nn
\eea
The condition $j\wedge j =\ft12 \Omega_2 \wedge \bar \Omega_2$ implies
\beq
e^{-2A} = \sqrt{\sigma_2 \tau_2 -\vec\beta_2  \cdot \vec \beta_2} \; .
\eeq
We recall that $\phi=2 A=-2D$.

The $T^4$ metric computed from \eqref{g4} and \eqref{g42} is
\beq
g_{mn} =e^{2A}\,  \left(
 \begin{matrix} 
      \tau_2 & -\beta_{2}^{(1)} & -\beta_{2}^{(2)} & 0 \\
     -\beta_{2}^{(1)} & \sigma_2 & 0 & \beta_{2}^{(2)}\\
      -\beta_{2}^{(2)} & 0 & \sigma_2 & -\beta_{2}^{(1)}\\
      0 &  \beta_{2}^{(2)} & -\beta_{2}^{(1)} & \tau_2 \\
   \end{matrix}
\right)   
\eeq
and the solution becomes
\bea
ds^2 &=e^{2A} \sum_{\mu=0}^3 dx^\mu dx_\mu + \sum_{m,n=1}^4 g_{m n} \d y^m  \d y^n  +  e^{-2A} | h (z) |^2 dz d\bar{z} \; ,  \nn\\
C_2 &=  \tau_1 dy^1 \w dy^4 + \sigma_1 dy^2 \w dy^3- \beta_1^{(1)}(dy^1 \w dy^3 + dy^2 \w dy^4)+\beta_1^{(2)} (dy^1 \w dy^2 - dy^3 \w dy^4 )~,\nn\\
  C_0&  = 0 \; , \;   C_2 = 0 \; , \;  C_4 =0  \; .
\eea

\subsection{Relations between local solutions}
\label{sec:udual}
In the preceding sections, we presented three types of supersymmetric local solutions to type IIB supergravity. 
For each class, we displayed an example of solutions on $T^4$ characterised by 4 holomorphic functions. These three solutions can be related to each other acting with T- and S- dualities.  
 
Under T-duality  along a direction $y$, the metric in the string frame and the NSNS/RR fields transform as \cite{b87,b88,Lunin:2001fv}:\footnote{Our conventions are such that $B \to -B$, $B' \to -B'$ with respect to \cite{Lunin:2001fv}.}
\bea
g'_{yy} &= \frac{1}{g_{yy}}, \quad e^{2 \phi '} = \frac{e^{2 \phi}}{g_{yy}}, \quad g'_{y m} = \frac{B_{y m}}{g_{yy}}, \quad B'_{y m}= \frac{g_{y m}}{g_{yy}} \nonumber \\
g'_{m n} &= g_{m n}-\frac{g_{m y}g_{n y} - B_{m y}B_{n y}}{g_{yy}}, \quad B'_{m n}= B_{m n}- \frac{B_{m y}g_{n y}-g_{m y}B_{n y}}{g_{yy}}\nn\\
 C'_{m ... n \alpha y} &=C_{m ... n \alpha} - \left( n - 1 \right) \frac{C_{[ m ... n | y}g_{y | \alpha ]}}{g_{yy}} \nonumber \\
C'_{m ... n \alpha \beta} &= C_{m... n \alpha \beta y} - nC_{[m ... n \alpha }B_{\beta ] y} - n(n-1)\frac{C_{[ m...n | y}B_{|\alpha | y}g_{|\beta ] y}}{g_{yy}}
\eea
 On the other hand, under S-duality for backgrounds with $C_0=0$ is
\beq
 \phi'= -\phi \qquad g'=e^{-\phi } g \qquad C_2' =- B \qquad B' = C_2~,  \label{sdual} 
 \eeq 
    
Using these formulas one can check that  solutions in sections  \ref{sexamplea}, \ref{sexampleb} and \ref{sexamplec} are related by the duality maps
 \beq\label{eq:dualities}
   A   \quad  \overset{  T_{14} }{\longleftrightarrow}   \quad C
 \quad  \overset{ S }{\longleftrightarrow}   \quad    B 
 \eeq
if $\beta^{(3)}$ is set to zero in section  \ref{sexamplea}.  It is important to notice that  unlike in the case of three holomorphic solutions studied in  the companion paper \cite{paper:global},
 solutions with four holomorphic functions cannot be map to purely metric backgrounds using dualities.  Indeed, a simple inspection of
 (\ref{bsol4}) shows that $B$ have legs long all 6 two-cycles of the $T^4$ so there is no way so translate it into metric via T-dualities.

\section{Conclusions and outlook }
\label{sec:conclusion}

  In this paper, we presented explicit solutions where the ten-dimensional spacetime takes the local form $\mathbb{R}^{1,3}\times M_4\times \Sigma$,   with $M_4$ a generalised complex manifold with $SU(2)$ structure and $\Sigma$ an open subset of $\mathbb C$. The metric, dilaton, NS and R potentials are oriented along $M_4$ 
   and assumed to vary only along $\mathbb{C}$.   We display explicit examples for $M_4=T^4$ specified by up to four holomorphic functions. These solutions can be viewed as supersymmetric solutions
    of ${\cal N}=(2,2)$ maximal supergravity in six dimensions with a set of scalar fields varying over the $z$-plane. 
   
    This theory is parametrised by a scalar manifold 
\be
{\cal M}_{IIB~{\rm on}~T^4}= SO(5,5,\mathbb Z)\backslash {SO(5,5,\mathbb{R} )\over SO(5,\mathbb R)\times SO(5,\mathbb R) } \;  \label{mt4} 
\eeq
of dimension $25$: $9$ fields parametrise the symmetric and traceless metric on $T^4$, $2 \times 6$ fields correspond to NSNS and RR two-forms and 4 fields are related to the dilaton, the $T^4$-volume, the RR zero- and four-forms. The  holomorphic functions $\varphi_I(z)$ characterising
the local solutions span a complex sub-manifold of (\ref{mt4}). For example, for solutions of class A  with metric conformally flat, the holomorphic functions  span the $n=1,\ldots 5$-complex dimensional submanifold  
   \beq
{\cal M}_{BPS}=  SO(2,n,\mathbb Z) \backslash{SO(2,n,\mathbb{R} )\over SO(2,\mathbb R)\times SO(n,\mathbb R) } \subset {\cal M}_{IIB~{\rm on}~T^4} \; .
 \label{mbps}
\eeq 
   Explicit solutions from class B and C with $n=1,\ldots 4$ were constructed in \ref{sexampleb}, \ref{sexamplec}. 
 They are U-dual versions of solutions in class A, and thus share the same moduli space (\ref{mbps}); in this case, the three solution classes correspond to different  orientations of   ${\cal M}_{BPS} $ inside ${\cal M}_{IIB~{\rm on}~T^4}$. 
 
The moduli space (\ref{mbps})   is isomorphic to the moduli space ${\cal M}_{K3,n }$ of complex structures for an algebraic $K3$ surface with Picard number $20-n$. The holomorphic functions characterising the flux solutions can then be viewed as  the complex structure of an auxiliary $K3$ surface varying holomorphically over a plane.  
A consistent fibration of the $K3$ surface then defines a fully consistent, non-perturbative flux solution of type IIB supergravity 
using results in \cite{Martucci:2012jk,Braun:2013yla,paper:global}.
In particular, a compact Calabi-Yau threefold composed of a $K3$ surface fibered over $\mathbb{CP}^1$ can be 
used to construct a four-dimensional vacuum of type IIB supergravity with non-trivial fluxes.
 In the global solution, the local solutions  are glued  together using U-dualities to cover the whole complex plane $\mathbb{C}$.
 Branes are associated to singular points in $\mathbb{CP}^1$ where the complex structure of the $K3$ fiber degenerates and around which the holomorphic functions defining the local solutions have non-trivial U-duality monodromies. For a critical number of branes, the two-dimensional metric can be chosen to be regular at infinity, thus compactifying the $\mathbb{C}$ plane into  $\mathbb{CP}^1$. In this procedure, fluxes translate into geometry and we can exploit the well developed techniques of algebraic geometry to find new  supersymmetric flux vacua.  The reader is referred to the accompanying paper \cite{paper:global}, where the details of the flux/geometry dictionary are studied and discussed in great detail.

\section*{Acknowledgements}

The authors would like to thank A.~Braun, V.~Braun, C.~Hull, L.~Martucci, M.~Petrini and D.~Waldram for interesting discussions and valuable comments. 
CD, ML and JFM would like to thank the Mathematical Institute, University of Oxford and Theoretical Physics group at Imperial College London for their kind hospitality during parts of this project. The work of PC is supported by EPSRC grant BKRWDM00.  AC would like to thank the University of Oxford and the STFC for support during part of the preparation of this paper. The research of ML was supported by the Swedish Research Council (VR) under the contract 623-2011-7205. The work of JFM is supported by EPSRC, grant numbers  EP/I01893X/1 and EP/K034456/1 and the ERC Advanced Grant n.~226455.  
 
 
\newpage  
\begin{appendix}

\section{Conventions}
\label{ap:conv}

 We use $(M,N,P,Q...)$ to index ten-dimensional quantities, $(m,n,p,q...)$ in the internal six dimensions and $(\mu, \nu, \rho,..)$ for the four space-time dimensions. Flat tangent space indices will sometimes be used, and we denote them with a hat: $\hat{M}, \hat{m}$ etc. 
 
 \vspace{12pt}
 \noindent
{\bfseries Gamma matrices.} Ten-dimensional Dirac matrices are denoted $\Gamma^M$, and six-dimensional ones $\gamma^m$. We will choose the latter to be  hermitian, $\gamma_{\hat{m}}^{\dagger} = \gamma_{\hat{m}}$,  imaginary and antisymmetric.  All Dirac matrices satisfy the Clifford algebra (e.g.~in six dimensions $\{\gamma_m,\gamma_n\} = 2 g_{mn}$)
and totally antisymmetric products of gamma matrices are denoted $\gamma_{m_1 m_2 ...m_k}$, where e.g.
\beq
\gamma_{m n} = \frac{1}{2} [ \gamma_m, \gamma_n] \; .
\eeq
The chirality operator in $d$ dimensions is given by 
\beq
\gamma_{d+1} = \ii^{-d/2} \gamma_{\hat{m}_1 ... \hat{m}_d} =\ii^{-d/2} \frac{1}{\sqrt{|g|}} \gamma_{m_1 ... m_d} \; ,
\eeq
where we use hatted letters for flat tangent space indices. The eigenvalues of $\gamma_{d+1}$ are +1 (-1) for chiral (antichiral) spinors. Thus, a six-dimensional spinor $\eta$ can be decomposed into chiral and anti-chiral components $\eta_{\pm}$, where $\gamma_7 \eta_{\pm} = \pm \eta_{\pm}$ ($\eta_-=\eta_+^*$).   Without loss of generality, we will take  $\eta_{\pm}$ to have unit norm, $\eta_{+}^{\dagger} \eta_{+} = \eta_{-}^{\dagger} \eta_{-} =  1$.

The two-dimensional Pauli matrices are given by
\beq
\sigma^1=\left(
\begin{array}{cc}
0 & 1 \\ 1 & 0 
\end{array}\right)
\qquad
\sigma^2=\left(\begin{array}{cc}0 & \ii \\ -\ii & 0\end{array}\right)
\qquad
\sigma^3=\left(\begin{array}{cc}1 & 0 \\ 0 & -1\end{array}\right) \; .
\eeq

\vspace{12pt}
\noindent
{\bfseries Differential forms and Hodge duals.}
We define the components of a differential $p$-form by
\beq
A = \frac{1}{p!} A_{m_1... m_p} d x^{m_1} \wedge ... \wedge d x^{m_p} \ ,
\eeq
The contraction of a $q$-form with a $p$-form ($p > q$) is 
\beq
B \lrcorner A = \frac{1}{(p-q)!} B^{m_1 ... m_q} A_{m_1 ... m_p} d x^{m_{q+1}} \wedge ... \wedge d x^{m_p} \ .
\eeq
Our convention for the Hodge star operation $*$, when acting on a $p$-form, is 
\beq
\label{eq:hodgedual}
* A = \frac{  \sqrt{|g|}}{p!(d-p)!} \epsilon_{m_{1} \ldots m_{d-p} } {}^{ n_1 \ldots n_p} A_{n_1 ... n_p}  d x^{m_1} \wedge ... \wedge d x^{m_{d-p} } \;.
\eeq
with $\epsilon_{1 \ldots d} = 1$, {\it{cf.}}~\cite{Andriot:2008va,Grana:2006kf}.  Another very useful relation is the combined identity
 \beq \label{eq:hodgewedge}
* (B \wedge A) = \frac{(-1)^{q}}{q!} B \lrcorner (* A) \ ,
\eeq
where $A$ is a $p$-form, $B$ a $q$-form, and $(d-p) > q$. Further relevant identities can be found in \cite{Gray:2012md}.

The six-dimensional exterior derivative can be decomposed into holomorphic and antiholomorphic parts, and with local holomorphic coordinates $z^a$, $a=1,2,3$, we have
\beq \label{eq:dc}
d=\partial+\bar{\partial}    \qquad  d^c ={\rm i}(\bar \partial-\partial) 
\eeq
with
\beq
\partial=d z^a {\partial \over \partial z^a}      \qquad \bar\partial =\d\bar z^a  {\partial \over \partial \bar z^a}  \, .
\eeq
The 2d Hodge dual satisfies
\beq \label{eq:star2}
*_{2}  d z ={\rm  i}\,  d z \; .      \quad\quad    *_{2}  \,1= d {\rm vol}_2 = \sqrt{|g_2|} \, d z \wedge \d\bar{z}  \; .
\eeq

\section{Spinors, structure groups and pure spinors} 
\label{app:purespinors}
   
\subsection{$SU(3)$ and $SU(2)$ structures}
 
\label{app:structures}

If a six-dimensional manifold admits a nowhere vanishing spinor $\eta_{\pm}$ its structure group is reduced to $SU(3)$. Another way to express this constraint is in terms of differential forms. The spinor can also be used to define a set of pure $O(6,6)$ spinors. In this appendix, we briefly review these different formalisms, in order to pave the way for the analysis of the local supersymmetry conditions of type~IIB compactifications.
  
  In string compactifications to four dimensions, supersymmetry requires the existence on $M_6$ of at least one globally defined and nowhere vanishing spinor $\eta$. The six-dimensional spinors $\eta_{\pm}$ can be used to build a nowhere vanishing real two-form $J$ and a complex decomposable three-form $\Omega_3$ on the six-dimensional manifold\footnote{Gamma matrices are in our conventions imaginary and complex.}
\beq
\label{eq:su3def}
 J_{mn}=-  \ii \eta_{+}^{\dagger} \gamma_{mn} \eta_{+}       \quad \quad\quad
\Omega_{mnp}  = - \ii \eta_{-}^{\dagger} \gamma_{mnp} \eta_+ 
\eeq
For manifolds of strict $SU(3)$ structure, i.e. those for which $\eta$ is unique, these are the only nowhere vanishing forms can be defined on $M_6$.\footnote{In particular, there are no globally defined one-forms on $M_6$; for spinors $\eta_{1,2}$ of the same chirality, bilinears $\eta_1^{\dagger} \gamma^{m_1...m_k} \eta_2$ vanish for odd $k$, and $\eta^{\dagger}_{\mp} \gamma^m \eta_{\pm}=\eta^T_{\pm} \gamma^m \eta_{\pm}=0$ follows from the antisymmetry of $\gamma^m$.}   
$J$ and $\Omega_3$  are subject to the constraints
\beq \label{eq:su3cond}
J \w \Omega_3 = 0 \quad \quad\quad
\frac{1}{6} \, J \w J \w J = \frac{ \ii}{8} \Omega_3 \w \bar{\Omega}_3=d {\rm vol}_6 \; .
\eeq
The Hodge duals of $J$ and $\Omega_3$ are 
\beq \label{eq:su3star}
*\!J = \frac{1}{2} J \w J \; \; \; \; , \; \; \; \;
* \Omega_3 = -\ii \Omega_3 \; .
\eeq
One can show that 
\beq
\label{eq:su3I}
{I^{\,p}}_n =  c\,  
 \epsilon^{p m_1\ldots   m_5}\,   ( {\rm Re}\, \Omega_3)_{ n m_1 m_2}\,  (  {\rm Re}\, \Omega_3)_{ m_3 m_4 m_5}  \; ,
\eeq
 satisfies $I_m{}^p I_p{}^n = -\delta_m^n$ for a given normalisation constant $c$. The matrix $I$ thus defines an almost complex structure \cite{Hitchin:2000jd} (see also sec. 3.1 in \cite{Larfors:2010wb}). Moreover, the contraction of $I$ with $J$ gives a metric
\bea
g_{mn}= -J_{mp} \, I^p{}_n   \; .
  \label{eq:su3met}
\eea
On a Calabi--Yau manifold, $J$ and $\Omega_3$ are closed and $I$ is an integrable complex structure. The Ricci-flat Calabi--Yau metric is given by \eqref{eq:su3met} once the K\"ahler form and holomorphic top-form have been correctly identified in the cohomology classes of $J$ and $\Omega_3$.

If a six-dimensional manifold allows two orthogonal nowhere-vanishing spinors, $\eta$ and $\chi$, its structure group is further reduced to $SU(2)$. Again, without loss of generality, we take the chiral and antichiral parts of the spinors to have unit norm.  The $SU(2)$ structure is characterised by the existence of a nowhere vanishing complex one-form $K$, a real two-form $j$, and a complex two-form $\Omega_2$ given by:
\vspace{-12pt}
\beq
\label{eq:su2def}
K_m =  \eta_{-}^{\dagger} \gamma_{m} \chi_+ \quad \quad
j_{mn} = - \ii \eta_{+}^{\dagger} \gamma_{mn} \eta_{+} +  \ii \chi_{+}^{\dagger} \gamma_{mn} \chi_{+}\quad\quad
\Omega_{2mn} =  \eta_{-}^{\dagger} \gamma_{mn} \chi_-
\eeq
The $SU(2)$ structure can be embedded into the $SU(3)$  via the relations 
\beq
\label{su2su3}
J = j + \frac{ \rm{i} }{2} K \w \bar{K}  \; \; \; \; , \; \; \; \;
\Omega_3 = K \w \Omega_2 \; .
\eeq
Using these relations and \eqref{eq:su3cond}, it is straightforward to show the $SU(2)$ structure relations 
\beq \label{eq:su2cond}
j \w \Omega_2 = 0 \; \; \; \; , \; \; \; \;
j \w j = \frac{1}{2} \Omega_2 \w \bar{\Omega}_2 \; \; \; \; , \; \; \; \;
K \lrcorner j = K \lrcorner \Omega_2 = 0  \; .
\eeq
 and vice versa these conditions imply that $J$ and $\Omega_3$ given by (\ref{su2su3}) is an $SU(3)$ structure (using the fact that $|K|^2 = 2$, which follows from the unit norm of $\chi_+$).

\subsection{Pure spinors} \label{ap:ps}

For manifolds with $SU(2)$ structure, one can write
\beq
\eta^1_+ =a \eta_+ \quad \quad\quad
\eta^2_+ = b(\cos \alpha \ \eta_+ +\sin \alpha \ \chi_+)\; ,
\eeq
 In presence of D-branes the modulus of the two spinor should match\footnote{This follows from the fact that D-brane boundary conditions relate left and right moving spinors. In particular for a Dp-brane one finds  $\epsilon_1=\hat \Gamma^{i_0 \ldots i_p} \epsilon_2$.}, and  supersymmetry requires
  $|a|^2=|b|^2=e^A$.  We write $a=|a| e^{\ii\theta_1}$, $b=|b| e^{\ii\theta_2}$. The parameter $\alpha$ interpolates between strict $SU(3)$ ($\alpha=0$, parallel spinors) and $SU(2)$ ($\alpha=\pi/2$, orthogonal spinors)  structures. The $O(3)$ spinors $\eta_{1,2}$ can be used to define two pure $O(6,6)$ spinors  
\beq
\Phi_{\pm} = \frac{1}{|a|^2} \eta^1_+ \eta^{2 \dagger}_{\pm}
= \frac{1}{8 |a|^2} \sum_{k=0}^6 \frac{1}{k!} \eta^{2 \dagger}_{\pm} \gamma_{m_k...m_1} \eta^1_+ \gamma^{m_1..m_k}
\label{pure}
\eeq
The right hand side of (\ref{pure}) can be thought as a polyform 
 via the Clifford map
\beq
  \gamma^{m_1 m_2 ... m_k} ~~\longleftrightarrow~~  d x^{m_1} \w d x^{m_2} \w ... \w d x^{m_k}
\eeq
 In particular bilinears made out of spinors of the same (different) chirality lead to odd (even) forms.  

The various contributions can be written as
\beq
\label{eq:su2alt}
\begin{split}
&\chi_+ = \frac{1}{2} K_m  \gamma^{m} \eta_- \\
&\eta_+ \eta_+^\dagger = \frac{1}{8} e^{-\ii j + \frac{1}{2} K \w \bar{K}}
\\
&\chi_- \eta_-^\dagger = \frac{1}{8} \Omega_2 \ e^{\frac{1}{2} K \w \bar{K}}
\; .
\end{split}
\eeq
 Equations (\ref{eq:su2alt})  are equivalent to (\ref{eq:su2def}) and can be used as an alternative definition of an $SU(2)$ structure.  
The equivalence between the two can be shown by multiplying the last two relations in \eqref{eq:su2alt}   by $({\mathbb{I}}, \gamma_m, \gamma_{mn},...,\gamma_{mnpqrs})$ and tracing over spinor indices; for the first relation one should multiply $\chi_+$  in \eqref{eq:su2alt} with $\eta_-^{\dagger} \gamma_n$ from the left. 

Plugging \eqref{eq:su2alt} into \eqref{pure} one finds
  \cite{Jeschek:2004wy,Andriot:2008va}  
\beq
\label{eq:purespin}
\begin{split}
\Phi_- &= -\frac{ e^{\ii \theta_+}}{8}  K \w \left(\sin \alpha \ e^{-\ii j} + \ii \cos \alpha \ \Omega_2 \right)\\
\Phi_+ &= \frac{e^{\ii \theta_-} }{8}  e^{\frac{1}{2}K\w \bar{K}} \left(\cos \alpha \ e^{-\ii j} - \ii \sin \alpha \ \Omega_2 \right) 
 \; ,
\end{split}
\eeq
 with $\theta_\pm=\theta_1\pm \theta_2$. The phase $\theta_+$ can be reabsorbed into the definition of $K$ so we discard this phase and rename $\theta_-=-\theta$ in the main text. By specifying $\alpha$ and $\theta$, we find different supersymmetric solutions, (see Section \ref{sec:local}).

\section{Explicit solution for the Killing spinor}
\label{sakilling}

In this appendix, we present explicit local type A solutions of the KSE \eqref{eq:Killing_2}-\eqref{eq:Killing_1} that are parametrised by up to five holomorphic functions.  It can be checked that this solution satisfies the equations of motion of type IIB supergravity, and we have done so using Mathematica. 

We start from the ansatz
 \bea
 \label{eq:ansatz5A}
d s^2& =e^{2A}\,  \sum_{\mu=0}^3 d x^{\mu} d x_{\mu} +e^{\phi-2A}\,  
\sum_{m,n=1}^4 \delta_{mn}\, dy^m\,  dy^n+   \, e^{2 D} \, |h(z)|^2 \, d z d \bar{z}   \nonumber\\
B& =b_a \,\chi_a^-   \qquad C_2= c_a \,\chi_a^-   \qquad   C_4=c \, d y^1\wedge dy^2 \wedge dy^3 \wedge dy^4
\eea
where we choose a basis of anti-self forms on $T^4$ as in (\ref{eq:chia})
\bea
\chi_a^- & =\{d y^1 \w d y^2 - d y^3 \w d y^4, d y^1 \w d y^3 + d y^2 \w d y^4,d y^1 \w d y^4 - d y^2 \w d y^3  \}
\eea
The KSE are then solved by Killing spinors $\epsilon^{i}$, that satisfy
\beq
 \label{eq:conditionA}
\epsilon^2={\rm i}\,  \epsilon^1  \qquad \Gamma^{z} \epsilon^1=0  \qquad \Gamma^{12} \epsilon^1= \Gamma^{34} \epsilon^1 \; ,
\eeq
with $\epsilon_0$ a constant spinor. The condition  $\Gamma^{12} \epsilon^1= \Gamma^{34} \epsilon^1$, together with the anti-self-duality of $B$ and $C_2$ implies that 
\beq
\label{eq:conditionB}
B_{mn} \Gamma^{mn} \epsilon =C_{mn} \Gamma^{mn} \epsilon=0 ~,
\eeq
where $\epsilon$ is the two-component vector $(\epsilon^1,\epsilon^2)^T$. The dilatino equation then reduces to the holomorphicity condition on the axio-dilaton field, 
\bea
\label{eq:KSE_cond0}
\bar  \partial  \tau =\bar \partial (C_0+{\rm i} \, e^{-\phi} )=0
\eea
in agreement with the pure spinor analysis in section \ref{sec:typeA}.

The vanishing of gravitino variations split into two conditions 
 \bea
 &\left( - H_{mn\bar z} \Gamma^{n \bar z} + \ft{\rm i}{2} e^\phi F_{ \bar z n p}\Gamma^{\bar z  n p} \Gamma_m 
\right)\epsilon^1 =0   \label{eq:KSE_cond1A}   \\
& \left( \nabla_m  -{\rm i} \frac{e^\phi}{8}\sum\limits_{n=1,5}\frac{1}{(n)!}F_{P_1 ... P_n}\Gamma^{P_1 ... P_n} \Gamma_m 
\right)\epsilon^1 =0\label{eq:KSE_cond1B}
\eea
Equation (\ref{eq:KSE_cond1A}) for $m = y^i$   give us
\beq
\left(- \bar\del B_{y^i n } \Gamma^{n \bar z} +\ft{\rm i}{2} e^\phi \left( \bar{\del}C_{np} - C_0 \bar{\del} B_{np} \right) \Gamma^{\bar{z}np}\Gamma_{y^i} \right) \epsilon^1= 0
\eeq
Writing 
\beq
\Gamma^{\bar z np} \Gamma_{y^i}=\{ \Gamma^{\bar z np} , \Gamma_{y^i} \} -\Gamma_{y^i} \, \Gamma^{\bar z np}=
2\Gamma^{\bar z [n} \delta^{p]}_i -\Gamma_{y^i} \, \Gamma^{\bar z np}   \label{gamidz}
\eeq
 and using  (\ref{eq:conditionB}) to discard the contribution of the last term in (\ref{gamidz}) one finds
\beq
\label{eq:KSE_cond2}
 {\rm i} e^\phi \bar{\del} \left( C_{n y^i} - \tau B_{ny^i} \right) \Gamma^{\bar{z}n}     \epsilon^1  =0
\eeq
that implies
\beq
\label{eq:parsolA}
\bar{\del} \left( C_{mn} - \tau  B_{mn} \right) = 0
\eeq
Let us consider now (\ref{eq:KSE_cond1B}),
\beq
\left( \frac{1}{4}\omega_{m n p}\Gamma^{n p}-{\rm i} \frac{e^{\phi}}{8}\left( F_n \Gamma^n +\ft{1}{5!}\, F_{mnopq} \Gamma^{mnopq} \right)\Gamma_m \right) \epsilon^1 =0     \label{condgrav}
\eeq
The non-trivial components of the spin connection are 
\bea
 \omega_{y^i y^i \bar{z}} &= \bar{\del}(e^{\phi -2 A}) \nn\\
 \omega_{z z \bar{z}} &=w^*_{\bar z \bar z z}= \ft12  \bar{\del} \ln {(\sigma_2 \tau_2 - \vec{\beta}_2^{2}) \over |h|^2}
 \eea
 Plugging this into (\ref{condgrav}), for $m=y^i$ one finds
\beq
\left( 2 {\rm i} \bar{\del}(e^{\phi -2 A})    +e^{2\phi-4A} (  \bar\del C_0 - e^{4A-2\phi} \, F_{\bar z 1234}   \right) 
\Gamma^{y^i \bar z} \epsilon^1=0 
\eeq
where we used $\Gamma^{1234} \epsilon^1=e^{4A-2\phi}  \epsilon^1$ and $\Gamma_{y^i}=e^{\phi-2A} \Gamma^{y^i}$. 
Writing $ \bar\del C_0 =-{\rm i}  \bar\del  e^{-\phi}$ one finds
\beq
 {\rm i} \,  \bar\del e^{\phi -4A}-F_{1234\bar z} =0 
\eeq
 or equivalently
 \bea
 \bar\del( C_4 - B \w C_2 + \frac{1}{2} \tau B \w B - \rm{i}  \, e^{\phi-4A} \, d^4y )=0.
 \eea
Thus, all three conditions in \eqref{eq:holpolyform} are reproduced.  

Finallly  taking $m=z$ in (\ref{eq:KSE_cond1B}) one finds the differential equation
\bea
\left( \del_{\bar{z}} - \frac{1}{8} \del_{\bar{z}} \ln \frac{\sigma_2 \tau_2 - \vec{\beta}_2^{2}}{|h|^2}  \right) \epsilon^1 = 0 ,\nonumber \\
\left( \del_z + \frac{1}{8} \del_{z} \ln \frac{\sigma_2 \tau_2 - \vec{\beta}_2^{2}}{|h|^2} \right) \epsilon^1 = 0 ,
\eea
where we use the compact notation $\vec{\beta}_2^2 = \sum_a (\beta^{(a)}_2)^2$. These equations are solved by
\bea
\epsilon^1= \left( \frac{\bar{h}(\bar{z})}{h(z) (\sigma_2 \tau_2 - \vec{\beta}_2^{2})^{1/2}} \right)^{1/4} \epsilon_0 ~.
\eea

\end{appendix}
 


\newpage
\bibliographystyle{JHEP}

\providecommand{\href}[2]{#2}\begingroup\raggedright\endgroup

\end{document}

%% file: mymacros.tex
\usepackage{amssymb, amsmath, amsthm,latexsym}
\usepackage[english]{babel} 
\usepackage[latin1]{inputenc} 
\usepackage{graphicx}
\usepackage[all]{xy}
\usepackage{bbm}
\usepackage{psfrag}

\newcommand{\be}{\begin{equation}}
\newcommand{\ee}{\end{equation}}
\def\bea#1\eea{\begin{align}#1\end{align}}
\newcommand{\nn}{\nonumber}

\newcommand{\w}{\wedge}

\renewcommand{\i}{\ensuremath{\textnormal{i}}}
\def\del {\partial}
\def\d {{\rm d}}

\newcommand{\cP}{\mathcal P}
